\documentclass[conference]{IEEEtran}
\IEEEoverridecommandlockouts
\usepackage{cite}
\usepackage{amsmath,amssymb,amsfonts}
\usepackage{algorithmic}
\usepackage{graphicx}
\usepackage{textcomp}
\usepackage{mathptmx}
\usepackage{epsfig, graphicx, caption,subcaption, float}
\usepackage{url,endnotes}
\usepackage{microtype}
\usepackage{latexsym}
\usepackage{url}
\usepackage{mdwlist}
\usepackage{pifont}
\usepackage{fixltx2e}
\usepackage{flushend}
\usepackage{pifont}
\usepackage{listings}
\usepackage[usenames,table]{xcolor}
\usepackage{amssymb,amsmath}
\usepackage{graphicx}
\usepackage{footnote}
\usepackage{tablefootnote}
\makesavenoteenv{tabular}
\makesavenoteenv{table}
\usepackage{wrapfig}
\usepackage{paralist}
\usepackage{multirow}
\usepackage{tabulary}
\usepackage{mathtools}
\usepackage{enumitem}
\usepackage{subcaption}
\usepackage{amssymb}
\usepackage{pifont}
\newcommand{\cmark}{\ding{51}}%
\newcommand{\xmark}{\ding{55}}%
\usepackage{array}
\usepackage[bottom]{footmisc}
\usepackage{enumitem}

\def\BibTeX{{\rm B\kern-.05em{\sc i\kern-.025em b}\kern-.08em
    T\kern-.1667em\lower.7ex\hbox{E}\kern-.125emX}}
    
\makeatletter
\newcommand{\linebreakand}{%
  \end{@IEEEauthorhalign}
  \hfill\mbox{}\par
  \mbox{}\hfill\begin{@IEEEauthorhalign}
}
\makeatother    
    
\begin{document}

\title{An Overview of UPnP-based IoT Security: Threats, Vulnerabilities, and Prospective Solutions}


\author{\IEEEauthorblockN{Golam Kayas}
\IEEEauthorblockA{\textit{Dept. of Computer \& Info. Science} \\
\textit{Temple University, USA}\\
golamkayas@temple.edu}
\and
\IEEEauthorblockN{Mahmud Hossain}
\IEEEauthorblockA{\textit{Dept. of Computer Science} \\
\textit{University of Alabama at Birmingham, USA}\\
mahmud@uab.edu}
\and
\IEEEauthorblockN{Jamie Payton}
\IEEEauthorblockA{\textit{Dept. of Computer \& Info. Science} \\
\textit{Temple University, USA}\\
payton@temple.edu}
\linebreakand 
\IEEEauthorblockN{ S. M. Riazul Islam}
\IEEEauthorblockA{\textit{Dept. of Computer Engineering} \\
\textit{Sejong University, South Korea}\\
riaz@sejong.ac.kr}

}
\IEEEoverridecommandlockouts
\IEEEpubid{\makebox[\columnwidth]{978-1-7281-8416-6/20/\$31.00~\copyright2020 IEEE \hfill} \hspace{\columnsep}\makebox[\columnwidth]{ }}

\maketitle

\IEEEpubidadjcol

\begin{abstract}
Advances in the development and increased availability of smart devices ranging from small sensors to complex cloud infrastructures as well as various networking technologies and communication protocols have supported the rapid expansion of Internet of Things deployments. The Universal Plug and Play (UPnP) protocol has been widely accepted and used in the IoT domain to support interactions among heterogeneous IoT devices, in part due to zero configuration implementation which makes it feasible for use in large-scale networks. The popularity and ubiquity of UPnP to support IoT systems necessitate an exploration of security risks associated with the use of the protocol for IoT deployments. In this work, we analyze security vulnerabilities of UPnP-based IoT systems and identify attack opportunities by the adversaries leveraging the vulnerabilities. Finally, we propose prospective solutions to secure UPnP-based IoT systems from adversarial operations.  
\end{abstract}


\begin{IEEEkeywords}
UPnP, IoT, Secure Service Discovery, Secure Service Advertisement, IoT Security
\end{IEEEkeywords}
\section{Introduction}
The emergence of the Internet of things (IoT) has introduced new opportunities for computing and communication. IoT devices have the potential to support smart applications across a wide variety of domains, such as smart cities, healthcare, manufacturing, and agriculture. As early deployments have shown positive outcomes and capabilities continue to advance, the use of IoT devices in modern infrastructures continue to be propelled forward; the number of Internet-connected IoT devices is projected to reach  24 billion by the end of 2020 \cite{numberOfIoT}.

Service-oriented architectures are well-suited to support IoT-enabled systems, providing the ability for resource-constrained IoT devices to advertise software services that can be discovered and used by other applications and services connected to the IoT network. The Universal Plug and Play (UPnP) has been widely embraced to support service-oriented IoT deployments. UPnP supports zero configuration dynamic discovery and advertisement of services, offering language independence and interoperability across heterogeneous devices, which allows for the creation of open, scalable IoT systems. 

However, a limitation in using the UPnP protocol to support deployments is that it was not designed to address security as a first-order concern. As a result, more than 20\% of UPnP-enabled products are exposed to known external and internal threats that take advantage of the UPnP protocol stack \cite{moore2013security, upnp-vul-note, khan2020reputation}. In addition, researchers have highlighted the need to address known implementation vulnerabilities in the UPnP software development kit~\cite{UplugUPnP}. Network  scanners like Shodan~\cite{Shodan} and ZMap~\cite{Zmap} reported millions of vulnerable IoT devices around the globe where enabling UPnP is the main cause of the vulnerability. Enabling unsecured UPnP to support applications across IoT networks can have severe consequences, as illustrated by the recent Mirai~\cite{Mirai}, Qbot~\cite{Qbot} and CallStranger \cite{callstranger} attacks.  


Securing the enormous number of devices that use UPnP is key challenge for security researchers and application developers \cite{alrawi2019sok}. Compounding the challenge is the fact that most IoT devices are battery-powered and have limited computational capabilities, making it difficult to simply adapt existing security solutions and to build secure models for UPnP-enabled IoT  devices. Identifying the key security vulnerabilities and threats for the use of UPnP in IoT networks and identifying potential solutions that consider the specific constraints of IoT systems in a systematic, unified review is an important first step toward addressing these gaps. In this paper, we present an analysis of the security vulnerabilities of UPnP service discovery, advertisement, eventing, and control methods in IoT networks. We also investigate the potential threat and the probable solutions. The contributions of this work are summarized as follows:
\begin{itemize}
    \item We analyze the vulnerabilities of different phases UPnP protocol.
    \item We identify the potential adversarial threats leveraging UPnP vulnerability.
    \item We present a review of the prior works attempting to secure UPnP protocol.
    \item We provide a security analysis supporting the proposed scheme can mitigate the security vulnerabilities and present a comparative discussion of the proposed model with prior works. 
\end{itemize}
\section{Background}
\begin{figure*}[t] 
 \includegraphics[width=\textwidth]{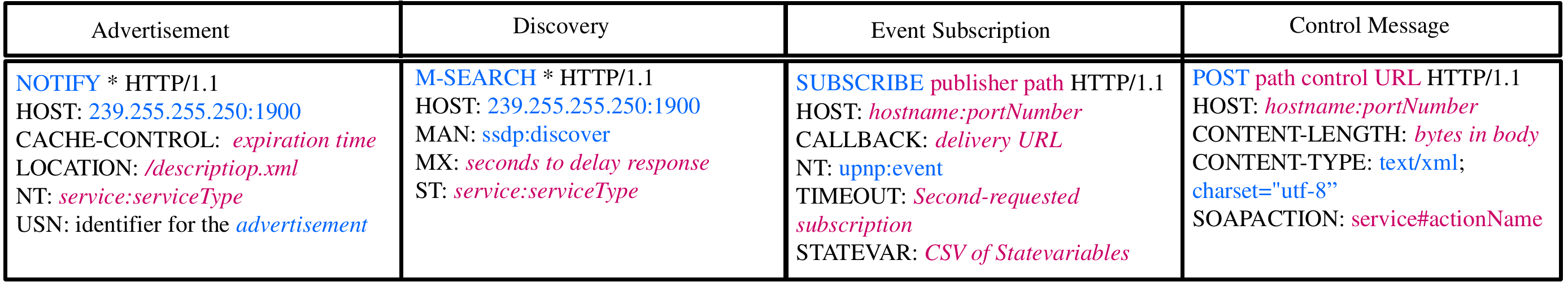}
  \begin{center}
  \caption{The messages used in UPnP interaction.}
  \label{fig:upnp-msg}
  \end{center}
\end{figure*}

\begin{figure}[t] 
 \includegraphics[width=\columnwidth]{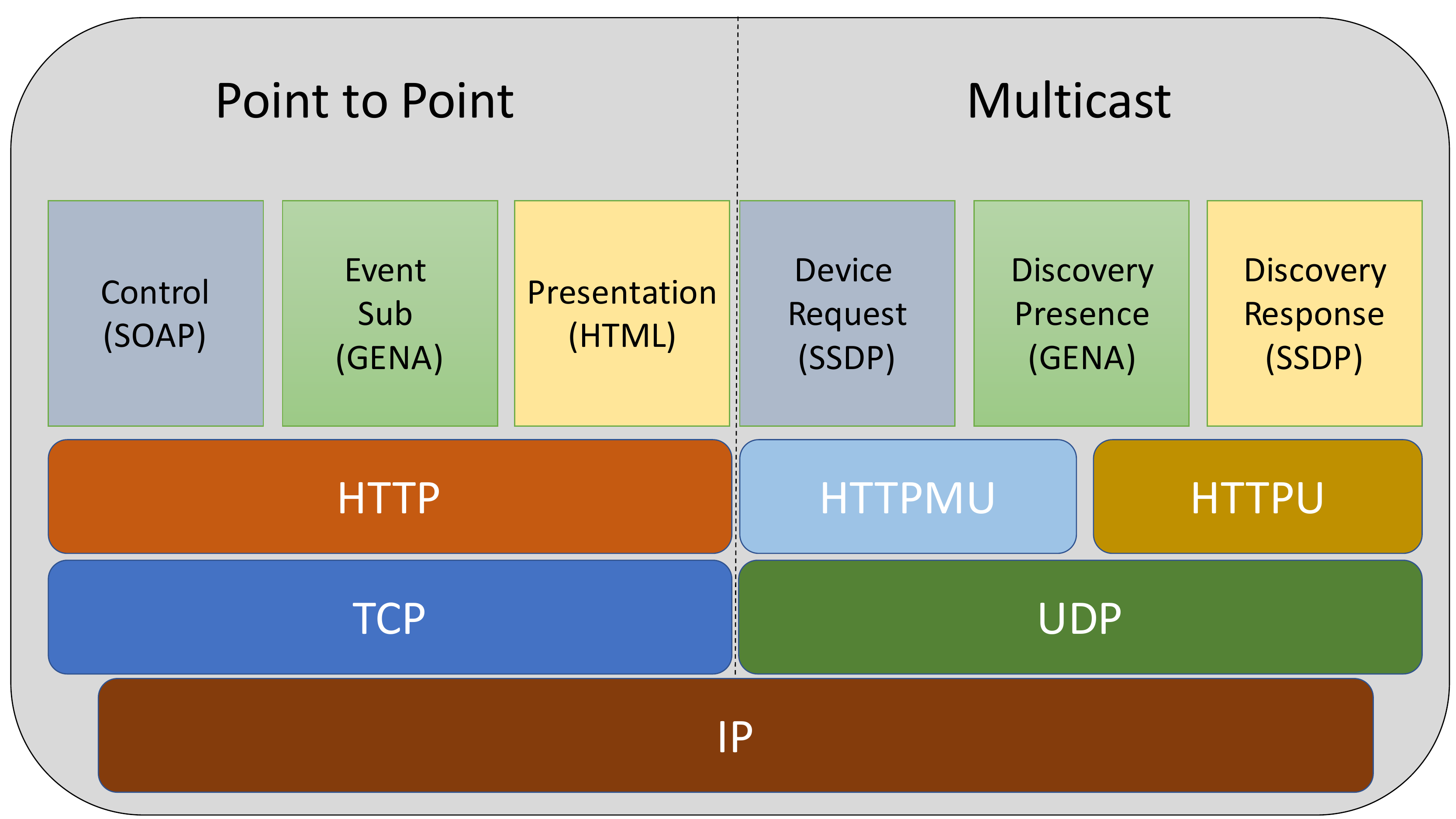}
  \begin{center}
   \caption{The UPnP stack.}
  \label{fig:upnp-stack}
  \end{center}
\end{figure}

In this section we present an overview of the UPnP protocol and its integration with IoT networks.

\subsection{UPnP Device Architecture and Interactions}
In this section, we present a brief description on the architecture of UPnP devices and their communication mechanisms. Universal Plug and Play (UPnP) leverages several popular protocols such as IP, TCP, UDP, HTTP,  XML or JSON to support seamless configuration and automatic discovery of services. Figure~\ref{fig:upnp-stack} shows the technologies used in the UPnP stack. 

\begin{figure}[t] 
 \includegraphics[width=\columnwidth]{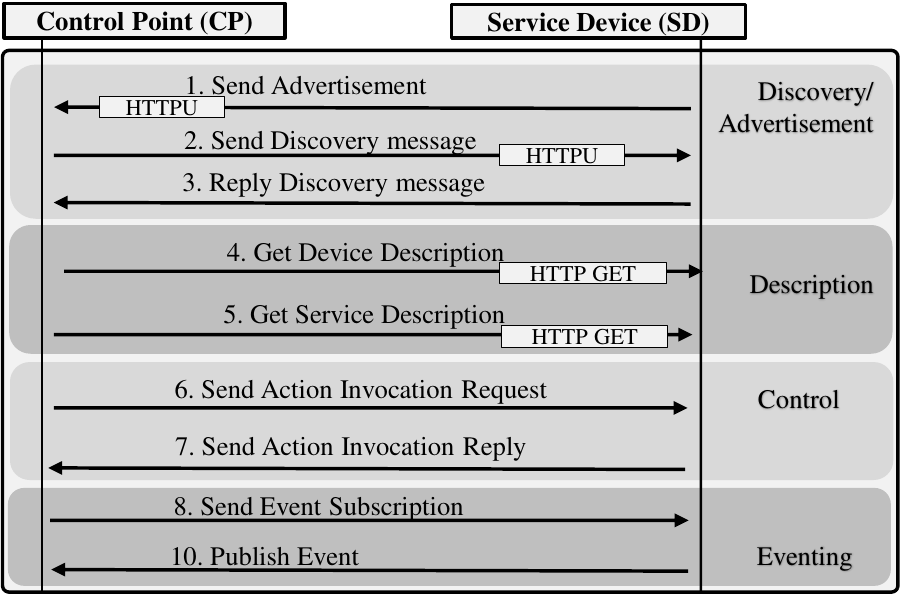}
  \begin{center}
   \caption{The Interactions in UPnP.}
  \label{fig:upnp-interaction}
  \end{center}
\end{figure}

UPnP devices are classified in two categories: service device (SD) and control points (CP). The service devices (SD) are basically servers that are responsible to delivering a service. The control points (CP) are the clients that consume the services provided by the SDs. A CP can be an application running on a device. For example, in a smart home the owner uses a smart thermostat that provide different services to control the temperature of the smart house. The  home owner communicates  with these services using an mobile application from his smart phone, where the application is refereed to as a CP in UPnP context. Figure~\ref{fig:upnp-interaction} shows the interaction between a CP and a SD in UPnP. The interactions are divided into three layers of UPnP stack, which are described in the following subsections.

\subsubsection{Discovery Layer} Discovery is the initial step of UPnP networking. This layer allows the SDs to advertise their services  and enables the CPs to search for a UPnP service in the network. UPnP devices use Simple Service Discovery Protocols (SSDP) for device discovery or advertisement. After joining the network an SD periodically sends advertisement messages to the network by multi-casting a \texttt{SSDP NOTIFY} message to a standard address and port (239.255.255.250:1900) as shown in Figure~\ref{fig:upnp-msg}. The \texttt{NT} field specify the service advertise by the \texttt{NOTIFY} message message  Simiarly, a CP  sends a multicast discovery request with \texttt{SSDP M-SEARCH} method on the reserved address and port (239.255.255.250:1900). The format of the discovery \texttt{M-SEARCH} message is shown in Figure~\ref{fig:upnp-msg}. The \texttt{ST} field mention the targeted service of the discovery message. Note that UPnP adopted HTTP over UDP protocol to send the discovery and advertise messages.

\subsubsection{Description  Layer} After discovering a service the CP has very little information about it. To use the service a CP needs to retrieve the description of the service and the device such as the attributes of the SD, the actions that can be invoked in the service, the state variables of the service, and couple of URLs to  send the action invocation request and to subscribe to a state change event of the service. The CP sends a \texttt{HTTP GET} request to the SD to retrieve the device description document. The device description exposes the physical and logical container of the SD such as the device serial number, the services provided by it. From the device description documents the CP gets a URL location to retrieve the service description document and sends another  \texttt{HTTP GET} to fetch the service description document.  The service description document defines actions that are accessible by a CP and their arguments. It also defines a list of state variables and their data type, range, and event characteristics. The state variable represents device state in a given time. Besides, the service description also provide two URLs for a service to invoke the actions and subscribe to a state change event of the service.

\subsubsection{Control \& Eventing  Layer} Once the CP has the information about the SD and its services, it can invoke actions from these services. To invoke an action, the CP sends a control message to the control URL of the service. Similarly, to track the state change of the service, the CP sends a event subscription message to the event URL. The event subscription message includes a \texttt{CALLBACL} URL (See Figure~\ref{fig:upnp-msg}), where the SD will publish the events.

\begin{figure}[t] 
 \includegraphics[width=\columnwidth]{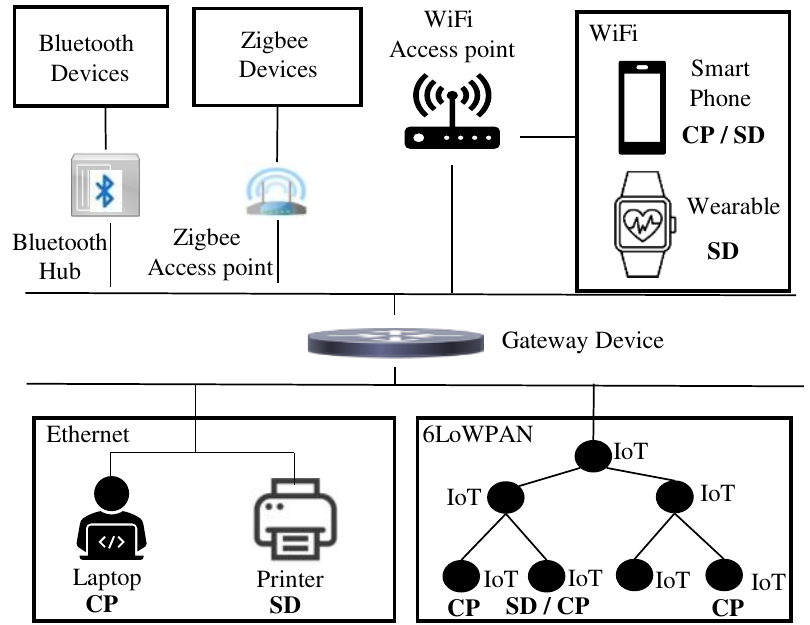}
  \begin{center}
   \caption{A UPnP-based IoT Ecosystem.}
  \label{fig:upnp-network}
  \end{center}
\end{figure}

\subsection{UPnP-based IoT Network}
Figure~\ref{fig:upnp-network} shows an example UPnP network with IoT devices. In a UPnP network, devices can be located in different types of networks such as BLE~\cite{bluetooth}, Zigbee~\cite{zigbee}, and 6LoWPAN~\cite{6lowpan}. The participants from different communication medium  interact with each other to perform UPnP operations. For example, a smart phone that uses WiFi can act as a CP, and can attempt to access a service provided by an IoT devices located in 6LoWPAN network. The gateway device is responsible in bridging different communication technologies.   
\section{Threat Model }
By design UPnP does not consider the security aspects of service discovery, advertisement, action invocation, and event subscription. In this section we discuss about the UPnP vulnerabilities and the identify the attacks take advantage of these vulnerabilities. We present a summary of the UPnP vulnerabilities and attacks based on the vulnerabilities in Table~\ref{tab:threat-model}. 
\begin{table*}[t]
	\centering
	\caption{UPnP Security Vulnerabilities and Attacks.}
	\label{tab:threat-model}
	\begin{tabular}{m{.20\textwidth}|m{.10\textwidth}|m{.10\textwidth}|m{.20\textwidth}|m{.20\textwidth}}
		\hline
		\multicolumn{1}{c|}{\textbf{Vulnerability}} &
		\multicolumn{1}{c|}{\textbf{Adversary}}  &  \multicolumn{1}{c|}{\textbf{Target}} & 
		\multicolumn{1}{c|}{\textbf{Attack}} &
		\multicolumn{1}{c}{\textbf{Impact}} \\ \hline \hline
         Absence of verification in Advertisement & Service Device &  Control Point & 
         \begin{itemize}
          \item Advertisement forgery
          \item Advertisement flooding
	    \end{itemize}
	    &	
	    \begin{itemize}
		    \item  Redirection to malicious URL
		    \item  Service impersonation
		    \item Resource Exhaustion
		\end{itemize}
		\\ \hline
		 Absence of verification is Discovery &  Control Point &  Service Device & 	
	 
         \begin{itemize}
          \item Discovery reply 
          \item Discovery flooding
          \item Discovery spoofing
	    \end{itemize}
	    &
		\begin{itemize}
		    \item Resource Exhaustion
		    \item Reflection and Amplification of malicious traffic
		    \item Distributed Denial of Service
		\end{itemize} 

		\\ \hline
	   Lack of authentication in Control &  Control point & Service device
	   &
	    Malicious Action invocation
	    &
	    	\begin{itemize}
		    \item Data leakage
		    \item Compromise network Security
		    \item Unauthorized access of data and service
		\end{itemize}
	     \\ \hline
	    Lack of integrity check in eventing  & Control Point & Service Device 
	    & 	
	    \begin{itemize}
		    \item Event Subscription forgery
		    \item Subscription flooding
		\end{itemize}
	    &
		\begin{itemize}
		    \item Resource Exhaustion
		    \item Reflection and Amplification of malicious traffic
		    \item Sensitive data leakage.
		\end{itemize}
		\\ \hline
	\end{tabular}
\end{table*}
\subsection{Vulnerabilities}
\subsubsection{No verification on service discovery and advertisement}. By default any participant of the UPnP network can advertise any UPnP services, similarly a discovery request to find an UPnP service can also be issued without any verification. There is no mechanism to verify whether a service device  is capable to provide the services, it is advertising. Likewise, a control point does not need to provide any proof of the capability to consume the service. 

\subsubsection{No access control on action invocation} A control point can invoke actions on UPnP services using the control messages. There is no access control policies to verify the credentials of the control point to invoke the requested actions on the service. The attackers leverage this issue and perform malicious operations on the service devices. 

\subsubsection{No verification on event subscription} A control point can issue an event subscription request to track state change of the services provided by a service device. The subscription request provides a \textbf{CALLBACK} URL where the state change events of the service will be published. The UPnP does not include any checks on the \textbf{CALLBACK} uses in publishing the events. The attackers take advantage of this lack of integrity check and used the event subscription mechanism in data ex-filtration and generating malicious traffic. 

\subsubsection{SDK vulnerabilities} There are many different portable, lightweight SDK implementation of UPnP  available for the IoT devices such as MiniUPnP \cite{MiniUPnP}, libUPnP \cite{libUPnP}. These SDKs are often have implementation issues providing opportunities to the attackers. Several vulnerabilities has been reported \cite{CVE-2019-14363, CVE-2020-15893, CVE-2020-24376,CVE-2020-5524} exploiting these SDK implementation issues. For example, in CVE-2020-15893 it is reported that it is possible for the attackers to injects commands using a crafted payload nto the \texttt{ST} field of the UPnP discovery message. The main solution of the SDK vulnerabilities is patching the affected IoT devices with regular updates. The update images of the IoT devices can be pass through some static and dynamic code analysis  tools like  AddressSanitizer~\cite{serebryany2012addresssanitizer}, Valgrind~\cite{nethercote2007valgrind}. If the exploit proof of concept (PoC) known, researcher suggested very quick context-aware patching mechanism~\cite{zeng2019heaptherapy} to prevent heap based memory vulnerabilities as reported in \cite{CVE-2018-16890, CVE-2016-8863}. Although, detecting problems in the  update artifacts and distributing the updates among the IoT devices is a different research subject itself.

\subsection{Attacks based on Vulnerable Advertisement}
\begin{figure}[t] 
 \includegraphics[width=\columnwidth]{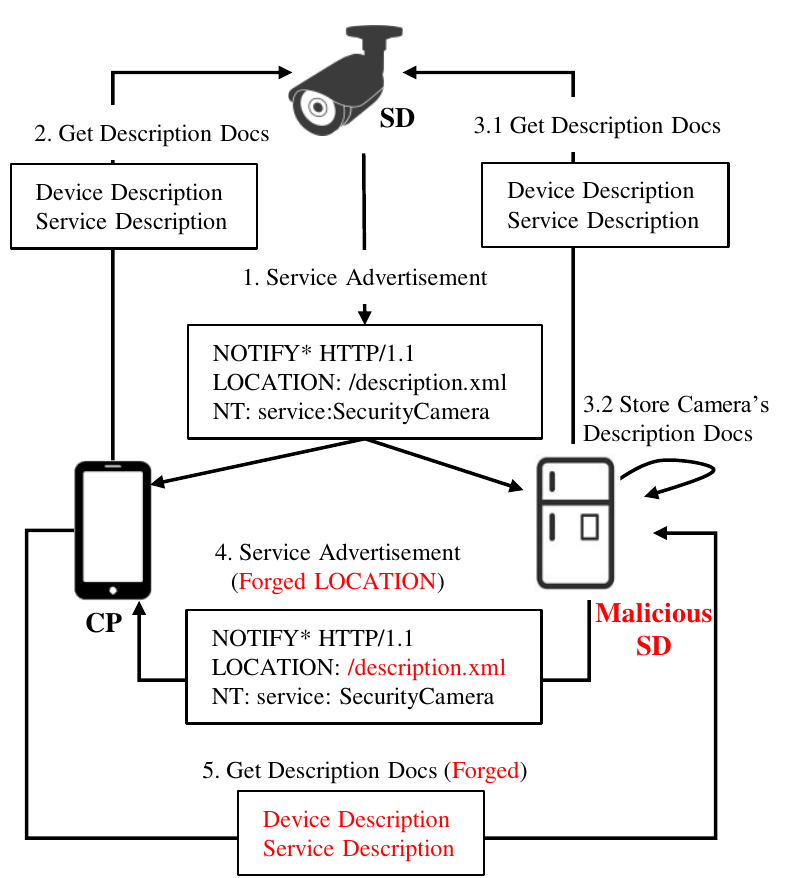}
  \begin{center}
  \caption{Service impersonation using Advertisement forgery.}
  \label{fig:svc-imp}
  \end{center}
\end{figure}
A malicious SD can use the leaked device and service description of a legitimate SD to advertise a service maliciously. As shown in Figure~\ref{fig:svc-imp} a security camera (legitimate SD) multicasts it's service \texttt{SecurityCamera} via advertisement message in the network. A legitimate CP (the smart phone) and a malicious CP (the refrigerator) receive the advertisement. The malicious CP retrieves the service description documents from the SD, stores the documents and craft a new advertisement message with a forged \texttt{LOCATION} URL of the description documents. Then the malicious SD multicast the forged advertisement in the network impersonating the \texttt{SecurityCamera} service. Thus the malicious SD (the refrigerator) tricks the CP ( smart phone) to believe that it provides the \texttt{SecurityCamera} service.  

\subsection{Attacks based on Vulnerable Discovery}
\begin{figure}[t] 
 \includegraphics[width=\columnwidth]{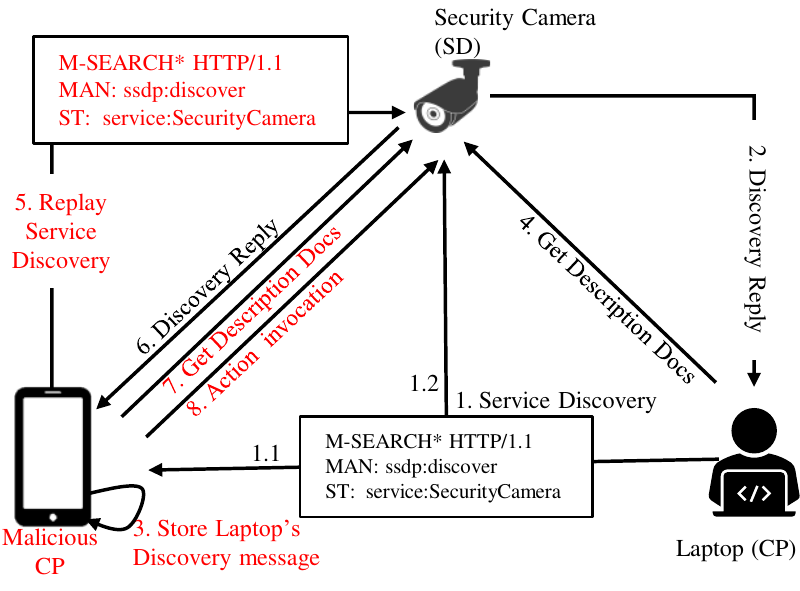}
  \begin{center}
  \caption{Discovery Message Reply by a Malicious CP.}
  \label{fig:dis-reply}
  \end{center}
\end{figure}

\noindent \smallskip \textbf{Discovery Message Reply: } In UPnP, any network participant can act as a CP by broadcasting a discovery message. As shown in Figure~\ref{fig:dis-reply}, when a legitimate CP (a laptop computer) broadcasts the discovery request searching a service \texttt{SecurityCamera}. An SD (security camera) and a malicious CP (smart phone)  receive the discovery request. The malicious CP stores the discovery request and replies it via network broadcast.

\begin{figure}[h] 
 \includegraphics[width=\columnwidth]{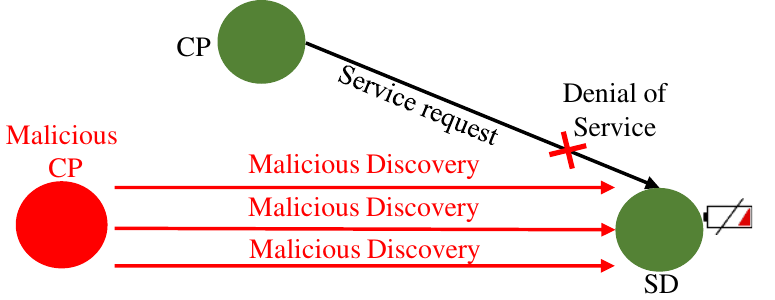}
  \begin{center}
  \caption{Denial of Service using Discovery Flooding.}
  \label{fig:dis-flood}
  \end{center}
\end{figure}

\noindent \smallskip \textbf{Denial of Service using Discovery Flooding: } The malicious CP floods a UPnP network by sending more discovery requests than usual. Thus a malicious CP can prevent a SD to provide important services to other legitimate CPs by keeping it busy, replying fake discovery message as shown in Figure~\ref{fig:dis-flood}. For example, an IoT-enabled pacemaker implanted in a patient’s body is supposed to send the heart-rate information frequently to a monitoring device. A malicious CP can flood the pacemaker device with fake discovery messages, compromising it's ability to report the heart-rate, causing a life threatening risk.     

\begin{figure}[t] 
 \includegraphics[width=\columnwidth]{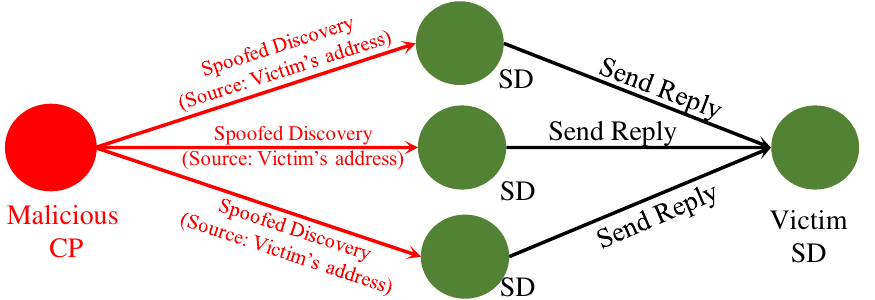}
  \begin{center}
  \caption{ Reflection and Amplification using spoofed Discovery.}
  \label{fig:dis-amp}
  \end{center}
\end{figure}

\noindent \smallskip \textbf{Reflection and Amplification using Spoofed Discovery: }  UPnP uses  HTTPU (HTTP over UDP) in the  discovery requests. Unlike TCP, UDP packets are vulnerable to source address spoofing. The malicious  CP  takes advantage of this fact and spoofs the  victim device’s IP address as the source of the discovery request as shown in Figure~\ref{fig:dis-amp}. As a result, an SD sends the discovery reply to victim devices. Moreover, the discovery request is received by all the SDs providing the service. So, all the SDs providing the targeted service will send discovery reply to the victim device. This spoofed discovery message attack  can generate huge malicious reflected and amplified traffic which is the building block Distributed Denial of Service (DDoS). A  study shows that spoofed discovery request has the potential to amplify the reflected traffic from a SD to target up to 30.8 times \cite{AmpUPnP}.

\begin{figure}[t] 
 \includegraphics[width=\columnwidth]{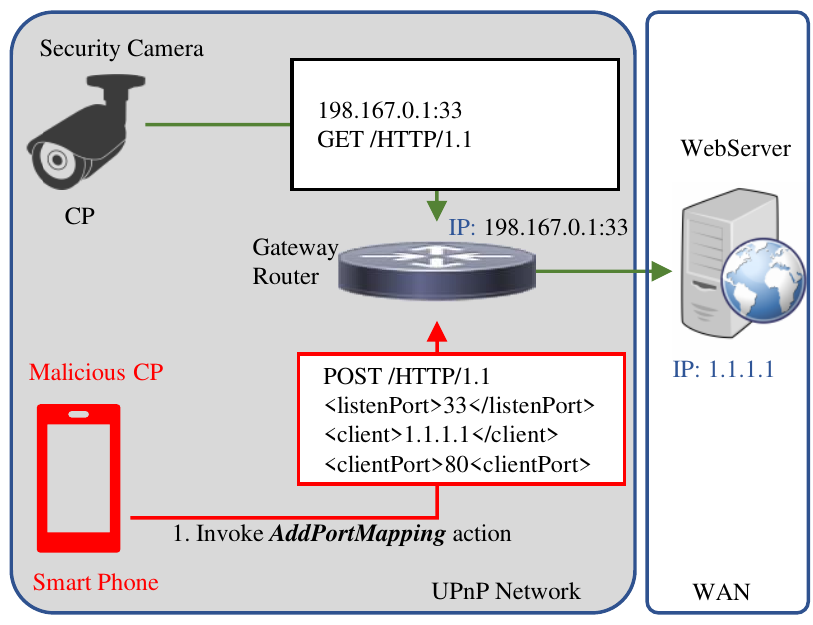}
  \begin{center}
  \caption{ The UPnProxy vulnerability exploiting UPnP action invocation.}
  \label{fig:upnproxy}
  \end{center}
\end{figure}
\subsection{Attacks based on Vulnerable Control}
The malicious CP can normally retrieve the description of the services from a SD. The service description includes the name of the actions can be invoked on a service, the required parameter and the URL to send the invocation request. After that CP can perform  action invocation request causing critical security issues. 
For example, the UPnP enable router often provide a service named  \texttt{WANIPConnections} with an action \texttt{AddPortMapping}. \texttt{AddPortMapping} is used to create a  port forwarding rule. A port forwarding rule is an application of network address translation (NAT) that redirects a communication from one address and port number combination to another while the messages are traversing a network gateway, such as an UPnP enabled router. Interestingly, a UPnP router is ofter connected to a outside network besides the UPnP enabled internal network. The adversaries can exploit that features of the UPnP enabled router and \texttt{AddPortMapping} to created unwanted proxies in the UPnP network. This attack is known as UPnProxy , and brought into the light by Akamai researchers \cite{upnp-proxy}. Figure~\ref{fig:upnproxy} shows an example of UPnProxy attack. In the figure, the smart phone acts as a malicious CP and invokes \texttt{AddPortMapping} to create a port forwarding rule, redirecting a traffic towards the gateway router to a web server external to the UPnP network. When a normal CP, the security camera makes a request to the specific port to the gateway router, using it's IP address, the gateway router redirects the request to the network external webserver which is not desirable to the security camera. Thus the malicious CP can setup a proxy on the gateway device to provide wrong information to the normal CP. Moreover this attack can be use to generate huge malicious traffic contributing in large scale DDoS attacks.

 \begin{figure}[t] 
 \includegraphics[width=\columnwidth]{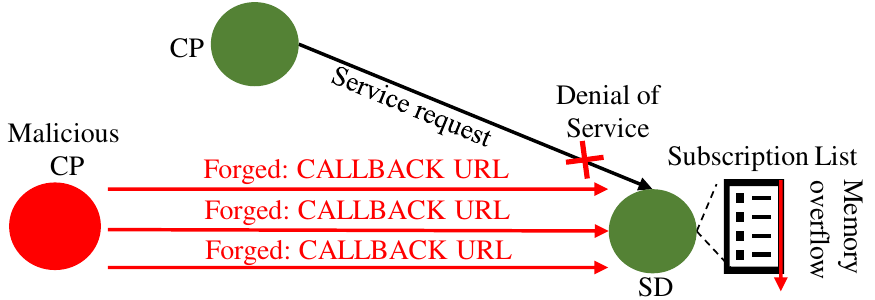}
  \begin{center}
  \caption{ Resource Exhaustion using Event Subscription request.}
  \label{fig:event-memory}
  \end{center}
\end{figure}

\begin{figure}[t]
  \begin{subfigure}[b]{\columnwidth}
    \begin{center}
    \includegraphics[width=\columnwidth]{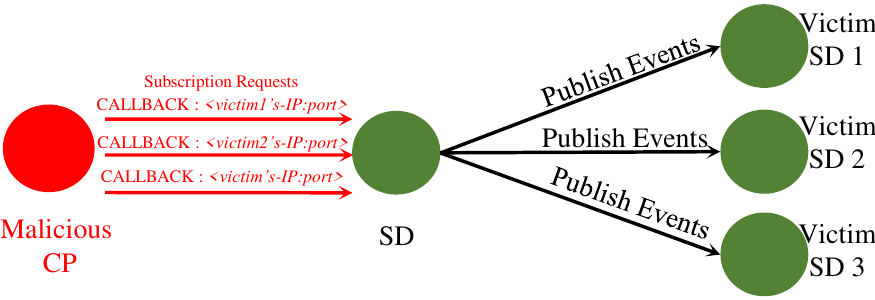}
  \end{center}
  \caption{Reflection using  Event Subscription.}
   
  \end{subfigure}
  \begin{subfigure}[b]{\columnwidth}
     \begin{center}
    \includegraphics[width=\columnwidth]{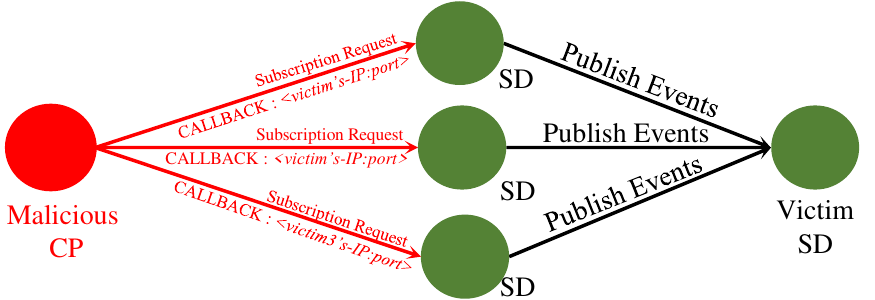}
  \end{center}
  \caption{Amplification  using  Event Subscription.}
   
  \end{subfigure}
  \caption{Reflection \&  amplification  using  Event Subscription CALLBACK forgery.}
  \label{fig:event-amp}
\end{figure}

\subsection{Attacks based on Vulnerable Eventing}
\noindent \smallskip \textbf{Resource Exhaustion using forged Event Subscription: } Figure~\ref{fig:event-memory} shows an memory exhaustion attack by a malicious CP using the event subscription flooding. A malicious CP can send a event subscription request to the SD. In response the SD stores a tuple of information to serve the subscription request: \texttt{<subscription-UUID (SID), callback-URL,timeout, http-version>}. According to the UPnP standard, the validity of the event subscription request (such as validity of the \texttt{CALLBACK URL}). The malicious CP takes advantage of this  and send huge number of forged event subscription requests with fake \texttt{CALLBACK URL}s. Eventually the SD will run out of memory storing information of the fake event subscription requests. Thus the attacker exhaust the resources of the SD, preventing it to serve other legitimate subscription requests.

\noindent \smallskip \textbf{Reflection and Amplification using Event Subscription: } A malicious CP can generate malicious traffic via reflections and amplification techniques by sending event subscription requests with a spoofed \texttt{CALLBACK} URL.
As shown, in Figure~\ref{fig:event-amp}, the CP uses  spoofed  \texttt{CALLBACK} URL in event subscription. As a result the SDs publish their events to the victim devices causing reflected and amplified traffic.  

\section{Limitations of The Related Work}
\label{sec:related-work}

In this section, we present prior efforts related to devising secure UPnP protocols.

KUPnP \cite{zhu2012kerberos} proposes an secure extension based on the Kerberos service \cite{kerberos} to protect the service devices and control points by introducing a key distribution center (KDC) as a central manager to handle authentication between devices and key management. However, the KDC is centrally managed to maintain a database consisting all the secret keys of the control points and service devices that results  scalability issues. Moreover, KUPnP provides no defense mechanism against impersonations attacks by an adversary tempering the service description document and removing the service required to perform Kerberos based authentication.

UPnP-UP \cite{UPnP-UP} is an extension that enables a multileyer security protocol. UPnP-UP can be used for user authorization and authentication  to achieve interoperability among the available services in a network. It provides a network manager to define access control policies using a well-defined user interface. Additionally, this provides the flexibility to select the security properties based on the need of UPnP networks (such as residence environment, commercial environment, secure environment).  Although, UPnP-UP focuses only on the service requests and action invocations which is associated with the eventing and control phase of UPnP. It does not provide any security mechanisms that address vulnerabilities in the discovery and advertisement phase of UPnP. Moreover, this scheme cannot verify the capability of the UPnP devices before joining the network, as SUPnP does. 

Guo et al. \cite{guo2013secure} proposes an UPnP key management scheme that is based on the group signature algorithm. The scheme includes the member join, signature verify and secure communication among the group members. But the assumption of Guo et al. is a small amount  devices need UPnP interconnections which is not suitable for IoT scenarios. If the group consists a large amount of IoT devices group signature based approach is not be scalable, thus cannot be deployed in large scale IoT networks. This work focuses only on the securing action invocation request, does not include securing discovery, advertisement and eventing.

U-PoT \cite{U-PoT} takes a different approach in securing UPnP. It uses honey pots to mitigate the attacks on the discovery and description phase of UPnP. In this approach, a honey pot is generated by automatically creating an emulated UPnP device from a UPnP device description document. While this approach may show promise in terms of detecting  malicious actions, event subscriptions, and replay attacks in the UPnP IoT network, it is expensive resource-wise and difficult to deploy. Moreover,  U-Pot does not stop  unauthorised discovery and action invocation of the UPnP services.


Islam et al \cite{islam2014user} proposed a simpler access control system without incorporating any complex authentication procedures by adopting access control list (ACL) to defend DLNA supported devices from unwanted control points. This work uses a particular field in requests or responses to identify a control point and manages a list of access rules. That approach is a good candidate to protect the UPnP devices from unauthorized control points due the light weight nature of the authentication procedures. But maintain a large number of access rules is always challenging, even not feasible in large IoT networks. This work focuses on the control phase of UPnP, and does not include the security concerns of discovery, advertisement and eventing phases. 

Vesa Pehkonen et al. \cite{pehkonen2010secure} proposes a secure UPnP network architecture using Transport Layer Security (TLS) to secure all TCP traffic, which carries most of UPnP discovery and advertisement messages. However, most IoT applications rely on UDP, as TCP is often considered to incur too much message overhead and energy consumption in networks of resource-constrained IoT devices. Besides, this work only focuses on the discovery architecture of UPnP, it does not provide any secure solution for UPnP control and eventing mechanism.


\section{Prospective Solutions}
\begin{figure}[t] 
 \includegraphics[width=\columnwidth]{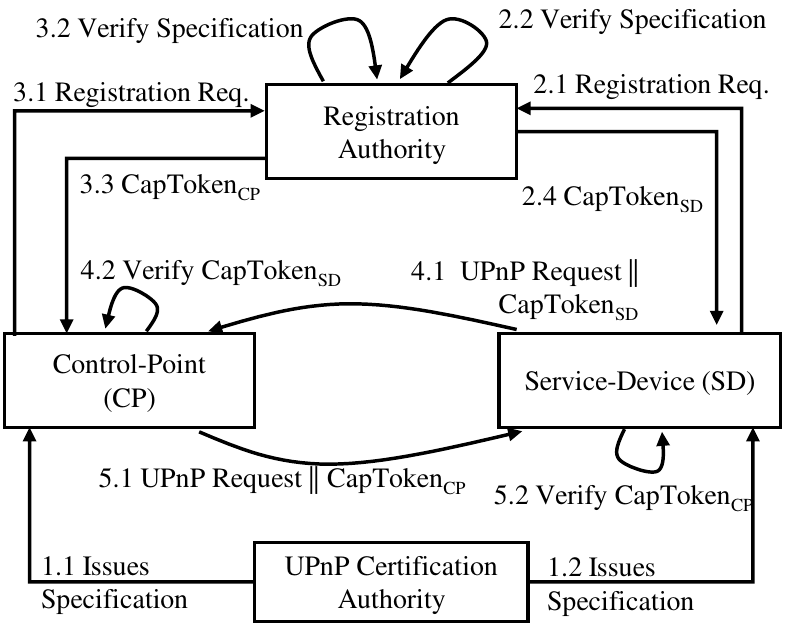}
  \begin{center}
  \caption{Overview of the proposed Secure UPnP protocol.}
  \label{fig:sys-overview}
  \end{center}
\end{figure}

In light of the security vulnerabilities and attacks identified, we propose a security scheme for UPnP that prevents malicious CPs to perform unauthorized operations and malicious SDs to advertise fake services. We propose a combine scheme of capability-based access control (CapBAC) and attribute-based access control (ABAC)  model to enforce authentication, verification, and access control in UPnP protocol.  In the CapBAC\cite{CapBac1,CapBac2,qiu2020survey} models , a user is provided a capability token (CapToken). The token contains the user's rights to perform certain operations on the services provided by an IoT device. A user sends its CapToken with a request. The device validates the CapToken to determine whether the user has permission to access the requested service. In ABAC \cite{ABAC}, the users are granted an operation based on their attributes, such as locations and time of the request. The ABAC model is accompanied by a list of rules or policies. The policies are invoked on the attributes, such as location and time of access, of the user to determine the user's access. A user is allowed to perform an operation if the policy invocation results in a permit decision. The policy invocation service considers various user's attributes to make a decision (permit or reject) on the operation request.

Figure~\ref{fig:sys-overview} shows the overview of our proposed model. In our proposed model, every participant of a UPnP network (both SD and CP) first go though an enrollment process. In the enrollment process, the participant receives a specification documents from a  trusted entity, Certification Authority. 
The role Certification Authority can be played by service vendor, device manufacturer, or the UPnP Certification Authority (UCA) \cite{upnp-certification}.
The specification documents state the hardware and software attributes of the participant. The certification authority signs the specification documents so that it can not be forged. The specification documents also contain a signature from the participant as a proof of the ownership of the document. This multi-signature feature, stops forgery and stealing of the specification document. The adversaries cannot forge the specification document because, adversary does not know the secret key of the  Certification Authority, so a forged specification will fail the verification process. Similarly, the adversaries can not use a leaked specification document as the document is signed by owner as well. 

After the enrollment process the participant is ready to join the UPnP network. In our proposed scheme, every UPnP network will implement a Registration authority (RA). Every participant sends a registration request to the RA before performing the UPnP operations. The registration request consists the specification documents of the participants and the UPnP actions the participant is interested in. 
RA verifies the specification documents from the trusted party to ensure the legitimacy of the participants hardware and software features mentioned in the specification document. If the specification document verification is passed, RA uses the hardware and software attributes of the participant to determine whether the actions requested by the participant is valid to perform. This that phase, RA deploys an attribute based access control scheme to validate the actions given the hardware and software attributes. After the attribute based policy invocation the RA assigns a CapToken to the participant. The CapToken is  cryptographically protected by the RA signature and determine the access right of the participant. Whenever the participant wants to perform an UPnP action in the network, it sends the CapToken with action request. The counter part verifies the CapToken with the help of RA and determine the access right of it's peer to perform the requested action. If the CapToken includes the permission to perform that operation, the participant gets the access to requested service or actions.        
\section{Security Analysis}

\begin{table*}[]
\centering
\caption{Security properties comparison of proposed scheme with related work.}
\label{tab:related-work}

\begin{tabular}{|p{0.15\textwidth}|p{0.05\textwidth}|p{0.055\textwidth}|p{0.05\textwidth}|p{0.055\textwidth}|p{0.05\textwidth}|p{0.055\textwidth}|p{0.05\textwidth}|m{0.055\textwidth}|}

\hline
\multicolumn{1}{|c|}{Scheme} & \multicolumn{2}{c|}{\begin{tabular}[c]{@{}c@{}}Malicious \\ Advertisement\end{tabular}} & \multicolumn{2}{c|}{\begin{tabular}[c]{@{}c@{}}Malicious\\ Discovery\end{tabular}} & \multicolumn{2}{c|}{Malicious Action} & \multicolumn{2}{c|}{\begin{tabular}[c]{@{}c@{}}Malicious Event\\ Subscription\end{tabular}} 
\\ \hline
        & Detection & Prevention & Detection & Prevention & Detection & Prevention & Detection  & Prevention 
        \\ \hline
UPnP   & \multicolumn{1}{c|}{\xmark}
       & \multicolumn{1}{c|}{\xmark} &  \multicolumn{1}{c|}{\xmark} &  \multicolumn{1}{c|}{\xmark}
       & \multicolumn{1}{c|}{\xmark} &  \multicolumn{1}{c|}{\xmark} & \multicolumn{1}{c|}{\xmark} 
       & \multicolumn{1}{c|}{\xmark}
\\ \hline
UPnP-UP \cite{UPnP-UP}  
        &\multicolumn{1}{c|}{\xmark}   
       & \multicolumn{1}{c|}{\xmark} &  \multicolumn{1}{c|}{\xmark} &  \multicolumn{1}{c|}{\xmark}
       & \multicolumn{1}{c|}{\cmark} &  \multicolumn{1}{c|}{\cmark} & \multicolumn{1}{c|}{\cmark} 
       & \multicolumn{1}{c|}{\cmark}
\\ \hline
U-PoT \cite{U-PoT}   &  \multicolumn{1}{c|}{\cmark}   
       & \multicolumn{1}{c|}{\xmark} &  \multicolumn{1}{c|}{\cmark} &  \multicolumn{1}{c|}{\xmark}
       & \multicolumn{1}{c|}{\cmark} &  \multicolumn{1}{c|}{\xmark} & \multicolumn{1}{c|}{\cmark} 
       & \multicolumn{1}{c|}{\xmark}
\\ \hline
KUPnP \cite{zhu2012kerberos} &  \multicolumn{1}{c|}{\xmark}   
       & \multicolumn{1}{c|}{\xmark} &  \multicolumn{1}{c|}{\cmark} &  \multicolumn{1}{c|}{\cmark}
       & \multicolumn{1}{c|}{\cmark} &  \multicolumn{1}{c|}{\cmark} & \multicolumn{1}{c|}{\cmark} 
       & \multicolumn{1}{c|}{\cmark} 
\\ \hline
Guo  et  al \cite{guo2013secure}    &  \multicolumn{1}{c|}{\xmark}   
       & \multicolumn{1}{c|}{\xmark} &  \multicolumn{1}{c|}{\xmark} &  \multicolumn{1}{c|}{\xmark}
       & \multicolumn{1}{c|}{\xmark} &  \multicolumn{1}{c|}{\xmark} & \multicolumn{1}{c|}{\cmark} 
       & \multicolumn{1}{c|}{\cmark}
\\ \hline
Pehkonen et al. \cite{pehkonen2010secure}  
            &  \multicolumn{1}{c|}{\cmark} & \multicolumn{1}{c|}{\cmark}
            &  \multicolumn{1}{c|}{\cmark} & \multicolumn{1}{c|}{\cmark}
            & \multicolumn{1}{c|}{\xmark}  & \multicolumn{1}{c|}{\xmark} 
            &  \multicolumn{1}{c|}{\xmark} & \multicolumn{1}{c|}{\xmark} 
\\ \hline

Islam et al \cite{islam2014user} 
            &  \multicolumn{1}{c|}{\xmark} & \multicolumn{1}{c|}{\xmark}
            &  \multicolumn{1}{c|}{\xmark} & \multicolumn{1}{c|}{\xmark}
            & \multicolumn{1}{c|}{\cmark}  & \multicolumn{1}{c|}{\cmark} 
            &  \multicolumn{1}{c|}{\xmark} & \multicolumn{1}{c|}{\xmark} 
\\ \hline
Proposed Scheme & \multicolumn{1}{c|}{\cmark}  & \multicolumn{1}{c|}{\cmark}
 & \multicolumn{1}{c|}{\cmark}  & \multicolumn{1}{c|}{\cmark}
  & \multicolumn{1}{c|}{\cmark}  & \multicolumn{1}{c|}{\cmark}
   & \multicolumn{1}{c|}{\cmark}  & \multicolumn{1}{c|}{\cmark}
\\ \hline


\end{tabular}
\end{table*}

In this section, we present a security analysis of the proposed scheme to show that it can prevent all the aforementioned UPnP vulnerabilities.

In the proposed scheme, every UPnP enable device or application need to present a certified specification document to register in an UPnP network. The specification documents are assigned by certified entity, certification authority (CA). The CA signs the documents as a proof of the legitimacy.  The specification documents also consist a signature by the owner of the documents as a proof of the ownership.  Thus, the adversaries are prevented to  enter into an UPnP network without proper specification.  Moreover, the multi-signed specification documents can not be stolen or leaked and can not be reused. Because, any leaked document fails the verification process of the registration authority (RA). This includes a challenge to solve a puzzle that requires the private key of the owner device, that used in assigning the ownership proof signature. Hence, the RA only pass the participant that provides legitimate specification. In the next stage the RA retrieves the hardware and software attributes oft he participant and apply an attribute based access control (ABAC) policy to determine the requested UPnP operations are compiled with the predetermined policy. If the request operations are permitted according to the ABAC policies given the hardware and software attributes, the RA assigns the access right as CapToken. This Captoken is presented with every UPnP operation request, that facilitates the receiving party to verify the authorization of the requesting party. For example, when a SD receives a discovery request it also receives a CapToken with it. The SD verifies the CapToken using the  signature of RA and from CapToken determines the capability of the CP to send a discovery message. Similarly, when a CP receives an advertisement or a discovery reply message, it verifies the permission of the SD to advertise the service using the CapToken. In case of the action invocation request and event subscription request the SD verifies the access right of the CP t perform the requested action. Hence, the proposed scheme enables verification in the both ends of service discovery and advertisement and ensure secure discovery and advertisement. Likewise, the propose scheme also enforce access control for service action invocation request and event subscription request using the CapToken.

Finally we present a comparison of the proposed scheme with prior work according to the security analysis provided in this section in Table~\ref{tab:related-work}.

\section{Conclusion}
In  this  work,  we  conducted  an  investigation on a popular protocol in service oriented IoT-ecosystem , UPnP.
Our investigation identify security vulnerabilities in service discovery, advertisement, control and eventing phases of UPnP. Our analysis highlighted how an adversary can launch service device impersonation, control point impersonation, denial of service, and resource exhaustion attacks when UPnP is deployed  in IoT  networks.
To  address  these  issues,  we introduced a capability based security  scheme,  which enables both service devices and control points to verify UPnP interactions which prevents the adversaries performing malicious activity. We provide a security analysis to show that proposed scheme successfully mitigate the vulnerabilities. We also provide a comparative analysis of the proposed scheme with prior works.
\section*{Acknowledgements}
This research was  supported in part by the US National Science Foundation (NSF) under Grant No. CNS-1828363 and in part the Sejong University research faculty program under the Grant No. 20192021.
\bibliographystyle{IEEEtran} 
\bibliography{bibliography.bib}
\end{document}